\documentstyle[preprint,aps]{revtex}
\def \lsim{ < \kern-10.5pt\lower+5pt\hbox{$\sim$}\kern+2.7pt}
\def \gsim{ > \kern-10.0pt\lower+5pt\hbox{$\sim$}\kern+2.7pt}
\begin{document}

\draft
% Assigned LF5803
%\twocolumn[ ....] will make one column for ... in twocolumn format
%\twocolumn[\hsize\textwidth\columnwidth\hsize\csname @twocolumnfalse\endcsname

\title{Direct Imaging of Submicron Scale Defect-induced Birefringence\\
in SrTiO$_{3}$ Bicrystals}

\author{E.~B.~McDaniel$\sp{(a)}$ and J.~W.~P. Hsu$\sp{(b)}$}

\address{
Department of Physics, University of Virginia,  
 McCormick Road, Charlottesville, Virginia 22901}

\date{\today}

\maketitle

\begin{abstract}
Using a near-field scanning optical microscope capable of quantitative polarimetry,
 we map the anisotropic strain fields associated with individual submicron defects 
near the fusion boundaries of SrTiO$_{3}$ bicrystals.  Many defects exhibit unexpected 
spiral-shape strain patterns, whose handedness is believed to be linked to the bicrystal 
synthesis process.  Direct observation of these defect-induced strain fields helps 
explain previously observed non-uniformity in the characteristics of high 
temperature superconductor grain boundary junctions fabricated on SrTiO$_{3}$ bicrystals.
\end{abstract}

\vskip2pc


%\vskip2pc]

A large amount of both experimental and theoretical work has been carried 
out in the attempt to achieve a thorough understanding of various 
mechanisms governing heteroepitaxial thin film growth.  
Of particular interest are inhomogeneous film growth systems, 
where the substrate itself is engineered to promote a desired lateral spatial pattern 
or feature\cite{QDs} in the film growth.  The general understanding of inhomogeneous 
growth currently lags well behind knowledge of simpler homogeneous systems.  
One crucial complicating factor that enters into inhomogeneous film growth 
is the effect of substrate-induced strain on the various physical properties of the films.

An important example of heteroepitaxial growth on engineered substrates is 
the growth of YBa$_{2}$Cu$_{3}$O$_{7}$ (YBCO) films on strontium titanate 
(SrTiO$_{3}$) 
bicrystals for the purpose of producing well-defined high-temperature 
superconducting (HTS) grain boundary Josephson junctions.\cite{dimos}  Bicrystals 
are made by fusing two pieces of single crystal together with a relative 
misalignment of their in-plane crystallographic axes.  The locations of 
the HTS Josephson junctions are then defined by the fusion boundary 
of the underlying SrTiO$_{3}$ bicrystal substrates.  In addition to the obvious 
technological applications, experimental results obtained on tricrystals made in 
a similar way have helped establish the fundamental symmetry of 
superconducting order parameters in this class of materials.\cite{tricrystal}  
Despite the wide use of bicrystals and tricrystals, some problems remain unresolved.
  For example, Josephson junctions fabricated on the same substrates often exhibit 
drastically different transport characteristics, and grain boundary 
inhomogeneities might disturb the overlap 
of superconducting wavefunctions.\cite{hilgenkamp}  
Therefore, a careful examination of how the substrate influences 
the YBCO film properties near the grain boundary, and hence device 
performance, is warranted.

Currently, most of the studies on this system have been either on transport 
property averaged over the entire junctions\cite{dimos,transport} or on microstructural 
characterization by transmission electron microscopy 
(TEM).\cite{dimos,TEM,TEMvoids}  
Spatial variations of transport characteristics and the relationship 
between structure and properties are obtained mostly through indirect 
inference or modeling.\cite{diffraction}  Only recently have direct measurements of 
spatial inhomogeneities of these HTS grain boundary Josephson junctions 
been examined.\cite{SEM}  In our previous work,\cite{YBCOAPL} 
we showed that the presence of 
submicron size defects near the SrTiO$_{3}$ bicrystal fusion boundary can 
cause severe boundary wandering in the epitaxial YBCO film and has a 
detrimental effect on the superconducting characteristics of the HTS 
Josephson junctions.  Due to the lack of surface topographic changes, 
we proposed that these defects cause strain in the surrounding substrate materials 
which in turn affect the YBCO thin film growth and physical properties.  
This study will focus on direct mapping of the strain fields associated 
with defects in SrTiO$_{3}$ bicrystals, which can then be used to predict 
{\it a priori} where YBCO film growth will be of poor quality.  Because these 
defects are typically of submicron dimensions, polarimetry measurements 
by conventional optical techniques are inadequate.
In this paper, we show how the sub-diffraction resolution of a near-field 
scanning optical microscope (NSOM) can be used with dynamic 
polarization-modulation (PM) techniques to achieve accurate, 
quantitative optical strain mapping with $\sim$ 100 nm spatial resolution 
non-destructively and quickly.

The bicrystals\cite{Nikko} consist of two pieces of (001) single crystal SrTiO$_{3}$, 
a cubic perovskite, fused together with a relative misalignment of the in-plane 
crystallographic axes.  The misalignments of all the bicrystals are symmetric, 
i.e., the [010] crystal direction is offset from the boundary direction by 
12$^{\circ}$ (18$^{\circ}$) 
on one side of the bicrystal and -12$^{\circ}$ (-18$^{\circ}$) on the other 
side for 24$^{\circ}$ (36$^{\circ}$) 
bicrystals.\cite{36deg}  The bicrystals are 1 cm x 1 cm x 0.5 mm and have been 
polished flat.  Transmission NSOM images of SrTiO$_{3}$ bicrystals reveal that 
the optical transmission along the fusion boundary is highly non-uniform (Fig. 1(a)), 
containing circular dark features of 0.1 to 1 $\mu$m in diameter.  Notice that 
there are no topographic changes associated with most defects (Fig. 1(b)).  
The optical contrasts\cite{deltaI} of these features vary from 4$\%$ to 67$\%$.  The NSOM 
resolution is required to observe these defects because when we imaged the same 
region of the sample with the tip pulled back from the surface by 
approximately one wavelength of the light ($\lambda$), 
the defects could not be resolved in this far-field 
image.\cite{MRSJulia}  Therefore, the defects seen in NSOM images must be located 
within $\lambda$ of the surface (i.e., in the near-field zone) and will thus 
have a large effect on epitaxial film growth.  These microstructural defects are 
probably due to sub-surface voids that have been reported in TEM studies 
of similar bicrystals.\cite{TEMvoids}  At the voids the local refractive index is effectively 
reduced from the bulk crystal and less light is collected due to refraction.\cite{NA}  
Consequently, the defects appear to be dark.\cite{index}

It is well known that if a solid contains regions of materials with different 
elastic moduli, strain fields are found in the regions of matrix material 
surrounding the inhomogeneities and a uniformly applied stress will be 
disturbed near the inhomogeneities.\cite{stresshole}  While the solution for ellipsoidal 
inhomogeneities in isotropic media is well known,\cite{eshelby} explicit calculations 
of strains in real crystalline solids can only be done by numerical methods\cite{shneck} 
because the elastic constants depend on crystallographic directions.  In the case of 
voids at the grain boundary of SrTiO$_{3}$ bicrystals, the calculation is 
further complicated because
elastic constants change abruptly at the grain boundary, in addition to the 
being different for the inhomogeneities and the matrix.  Furthermore, one 
need to consider the interactions between nearby inhomogeneities.  
Currently, there are no published theoretical calculations for strain fields 
associated with defects at internal grain boundaries, such as the case we
report here.  Thus, direct experimental measurements are needed to advance the
current knowledge in this field.

As a cubic perovskite crystal ($m3m$), SrTiO$_{3}$ is optically isotropic 
with negligible absorption in the visible range.  Presence of strain fields will, 
however, change the refractive index.  Thus, a non-zero birefringence, 
the difference in refractive index ($\Delta n$) experienced by the two 
eigen-polarizations, is a quantitative measure of local anisotropic strain fields.  
The NSOM experiments were done in air at room temperature with an instrument 
capable of operation both in transmission mode\cite{LAO} and in linear 
birefringence imaging mode.\cite{oakberg}  Dynamic birefringence measurements are 
performed by incorporating a photoelastic modulator (PEM) in the optical path 
to vary the polarization state of the light at 50 kHz.\cite{expsetup}  Detailed description on 
the design and performance of the polarization modulation NSOM has been 
reported in a separate publication.\cite{PMNSOM}  The physical quantity measured in this 
experiment is the linear retardance, which is the phase difference 
acquired by the two orthogonal, linear eigen-polarizations media after 
raveling through an anisotropic media.  Retardance and birefringence are 
linearly related.  The linear retardance (birefringence) information is 
contained in the signal components at the fundamental PEM modulation 
frequency ($1f$) and at the second harmonic ($2f$), while the $DC$ 
component is directly proportional to the transmittance of the sample and 
the laser intensity.  The dependence of the $1f$ and $2f$ components on sample 
transmittance and overall intensity fluctuations can therefore be removed 
by dividing out the $DC$ signal.  To obtain both the magnitude and the 
orientation of the local sample linear retardance, the demodulated 
$1f/DC$ and $2f/DC$ signals are recorded simultaneously at the same 
sample position.  As the sample is moved relative to the tip, images 
based on each of these signals are constructed. 
The linear retardance 
magnitude ($\phi$) and relative optic axis orientation ($\theta$) of the 
sample at a given point can be calculated from the simultaneously 
recorded $1f/DC$ and $2f/DC$ data according to\cite{PMNSOM}
$$
\phi = \sin^{-1} \sqrt{\frac{(1f/DC)^2} {(2J_1(\delta_0))^2} 
+ \frac{(2f/DC)^2} {(2J_2(\delta_0))^2}}
$$
and 
$$
\theta = \tan^{-1} [\frac{(2f/DC) J_1(\delta_0)} {-(1f/DC)J_2(\delta_0)}],
$$
where $J_1$ and $J_2$ are the first and second order Bessel functions, $\delta_0$
is the amplitude of the PEM modulation, $\phi$ is the magnitude of 
the retardance of the sample, and $\theta$ is the relative orientation of 
the local optical axis of the sample.  A PEM modulation amplitude $\delta_0$ 
of 2.405 radians (the first zero of $J_0$) was chosen so that the $DC$ signal 
is independent of sample retardance.\cite{gupta}  To accurately measure 
linear retardance quantitatively, we developed a method to prevent the 
unwanted retardance due to the fiber and the tip from contaminating the 
sample retardance.\cite{PMNSOM}  Before the sample was inserted for measurements, 
the linear retardance of the system is nulled to the noise limit.

The transmission NSOM image of an individual defect in a 24$^{\circ}$ bicrystal 
is shown in Fig. 2(a), while Fig. 2(b) shows the corresponding $\phi$ image.  
Linecuts across the two images are shown in Fig. 2(c) and 2(d), respectively.  
In the $\phi$ images, the grayscale ranges from zero (black) to 0.05 radian (white).  
Recall that $\phi$ is the magnitude of the total linear retardance and therefore 
has only positive values.  
$\phi$ is related to $\Delta n$
by
$\Delta n = \lambda \phi / 2 \pi d$,
where $d$ is the distance over which the change in phase is acquired.  
The strain $\epsilon$ associated with $\Delta n$ for a cubic material is given by
$\epsilon = 2 \Delta n / n^{3} (\Delta p)$,\cite{nye} 
where $n$ = 2.376 for SrTiO$_{3}$ at 670 nm\cite{gervais} and $\Delta p$ is related to 
photoelastic constants for SrTiO$_{3}$.  When the strain is compressive or 
tensile along the [100] or [010] crystallographic directions, $\Delta p = p_{11} - p_{12}$
which has the value of 0.144 for SrTiO$_{3}$.\cite{reintjes}  For compressive or tensile 
strains along the [110] directions or shear strains, $\Delta p$ 
is replaced by\cite{nye,reintjes} 
2$p_{44}$ = 0.055.  To calculate other strains, one must separate them into these 
components.  Since we do not induce the strain externally, we take 
$\Delta p \sim p_{avg} \sim$ 0.10 when extracting strain values 
from measured birefringence values.

As discussed above, in SrTiO$_{3}$ a non-zero $\phi$ indicates the presence of 
local anisotropic strain.  We found that the lateral dimensions of the 
retardance patterns are larger than the physical defect sizes appearing 
in the transmission images, as expected because strain fields propagate 
beyond the defects themselves.  From Fig. 2(d), it is evident that at the 
location of the defect, $\phi$ does not decay monotonically away from 
boundary, but oscillates.  The low $\phi$ regions between two high $\phi$ regions 
are nodal planes (lines) between a compressive and a tensile region.  
This non-monotonic dependence can only result 
from multiple sources of strain.  In this case, strains from both the 
void and the fusion boundary contribute.  When linecuts perpendicular 
to the boundary are taken well away from the defects, we find that $\phi$ does 
decay monotonically from a non-zero value at the fusion boundary to the 
background value, indicating further support that the oscillations in $\phi$ near 
the defects are the result of both voids and boundary.  The 
unexpected symmetry-breaking spiral pattern shown in Fig. 2(b) was 
found to accompany many, though not all, defects on the 24$^{\circ}$ 
bicrystals.  We discuss the possible origins later in the paper.  
The retardance patterns of the same defect are reproducible when 
imaged with different tips.

As seen in Fig. 2(c) and 2(d), maximum values of optical contrast 
and $\phi$ associated with each defect can be obtained despite 
the complex patterns in the $\phi$ images.  We found that 
defects with greater optical contrast in the transmission 
NSOM images usually have greater $\phi$ associated with them.  
The plot in Fig. 3 shows maximum $\phi$ vs. maximum optical contrast 
for 15 defects from three 24$^{\circ}$ bicrystals and 
6 defects from a 36$^{\circ}$ bicrystal.  Defects on the 24$^{\circ}$ bicrystal 
$\#$1 
(from an older batch) showed, on average, higher optical contrast 
and higher maximum $\phi$ than on the 24$^{\circ}$ bicrystals $\#$2 and $\#$3.  
Defects on the 36$^{\circ}$ bicrystal, however, showed higher optical contrast and 
maximum $\phi$ than any of the 24$^{\circ}$ bicrystals.  
The optical contrast ranges from 4\% to 23\% for defects on 
the 24$^{\circ}$ bicrystals, and from 24\% to 67\% for defects on 
the 36$^{\circ}$ bicrystal.  The maximum $\phi$ for defects on the 
24$^{\circ}$ bicrystals included in the plot ranges from 0.016 radians to 
0.087 radians, and ranges from 0.030 radians to 0.22 radians for 
defects on the 36$^{\circ}$ bicrystal.  The plotted points are also distinguished 
based on the particular objective used for light collection.  
Small systematic differences were noted in the maximum$\phi$
 when the same defects were imaged using different collection 
optics, with slightly higher $\phi$ values measured with lower numerical 
aperture (NA) collection objectives.\cite{objective}  Also of note in Fig. 3 are the 
two points which represent same defect on 24$^{\circ}$ bicrystal $\#$2 imaged 
with different tips.  The difference in these two measurements, 
$\leq$ 10 \%, is an estimate of the accuracy in our determination of $\phi$.

Despite this general trend in the plot, it is not possible to predict 
the linear retardance based only on the $DC$ optical contrast, as evidenced 
by the scatter.  Fig. 4(a) shows two defects on the 36$^{\circ}$ bicrystal which are 
similar in size and optical contrast, while Fig. 4(b) shows that one has a 
much stronger signature in the $\phi$ image.  The optical 
contrast is 40\% for the lower and 33\% for the upper defect.  
However, the maximum $\phi$ is 0.084 radians for the 
lower and almost none for the upper defect.  Since the density of defects on 
the 36$^{\circ}$ bicrystal is substantially higher than on the 
24$^{\circ}$ bicrystals,\cite{MRSJulia} the 
strain fields from neighboring defects overlap each other on the 36$^{\circ}$ bicrystal, 
as can be seen in Fig. 4(b).  Since 36.8$^{\circ}$ is a special grain 
boundary consisting mainly of symmetric segments,\cite{TEMvoids} 
one might expect the 
36.8$^{\circ}$ boundary to contain fewer defects.  
Our results show the contrary.  Since the 36$^{\circ}$ bicrystal we studied is from an 
older batch, these defects are probably process related.

When obtaining $\Delta n$ from $\phi$, there is a difference between 
NSOM and conventional far-field measurements. Though the 
light travels through the entire thickness of the sample, only features 
within approximately $\lambda$ of the surface can cause changes over the 
small scan size of NSOM images.  The bulk of the crystal contributes 
an overall background, and features beyond $\lambda$ can produce slowly 
varying features only.\cite{PMNSOM}  The retardance patterns associated with 
these defects, however, have features that are $\sim$ 100 nm.  Therefore, 
they must be due to strain fields near the surface.  For these features 
of interest, then, we take $d \sim \lambda$.  This results in a maximum $\Delta n$ of .014 
for the defects on the 24$^{\circ}$ bicrystals and .035 for defects on the 
36$^{\circ}$ bicrystal.  These birefringence values correspond to maximum strain of 
0.021 for defects on the 24$^{\circ}$ bicrystals and of 0.052 for defects on the 
36$^{\circ}$ bicrystal.

The spiral shape of the strain patterns is unexpected, in that it 
breaks the reflection symmetry of the boundary.  Inclusions of 
different thermal contraction constants or lattice constants in a 
bulk cubic crystal results in two- or four-fold symmetric patterns\cite{shu} 
and not the symmetry-breaking strain patterns observed here.  However, 
the defects we study in this paper are not solid inhomogeneities, but 
are voids that were formed at high temperatures during the fusion process.  
The surface tension of the surrounding SrTiO$_{3}$ surfaces plays a role in 
determining residual strain when the samples are cooled down.  The 
presence of a bicrystal boundary makes the problem even more complicated 
because the elastic constants change abruptly at the boundary due to 
a change in the crystallographic direction.  Furthermore, using NSOM we 
are probing near-surface defects.  The surface may allow the strain from the 
defect to relax.  In addition, the physical and chemical properties of the 
surface region might be different from the bulk.  Non-cubic strontium and 
titanium rich phases near the surface have been observed to form 
at the elevated temperatures.\cite{szot}  We also found that defects with clear spiral 
patterns in the same batch of 24$^{\circ}$ samples all show 
the same handedness.  This implies a 
connection to some macroscopic phenomenon, such as a twisting motion between 
the two halves of bicrystals that happens after the fusion process has initiated.\cite{takahashi}  
Despite the spiral shape, the observed retardance patterns are not due to screw 
dislocations.  Modeling the sample as a cubic crystal with no boundary, our 
calculations show that the spiral shape cannot be explained by the birefringence 
associated with the strain field from a dislocation\cite{weertman} in the bulk.  
Furthermore, TEM results do not show screw dislocations in SrTiO$_{3}$ 
bicrystals of such large misalignment angles.

In summary, we report a direct and quantitative 
measurement of anisotropic strain fields associated 
with near-surface defects at the fusion boundary of 
SrTiO$_{3}$ bicrystals with 100 nm spatial resolution.  
The results reveal unexpected strain patterns that we 
argue are linked to problems in the fusion process.  The combination 
of dynamic polarimetry and NSOM leads to high sensitivity and 
submicron resolution.  This technique will be useful in studying how 
strain from substrates or engineered patterns influences film growth 
and physical properties in a broad range of heteroepitaxial growth systems.

We thank J. Z. Sun for two of the bicrystals, D. Vanderbilt, 
J. Mitchell, R. Shneck, M. Kawasaki, and K. Takahashi for helpful 
discussions.  E. B. McDaniel acknowledges the support of an ONR PhD 
fellowship and J. W. P. Hsu of a Sloan Foundation Research Fellowship.  
This work is supported by NSF DMR-9357444.

\clearpage

\noindent {\bf Figure Captions}:\\
%\begin{figure}
%{\includegraphics[scale=0.6]{zero.ps}}
%\centerline{\epsfxsize=5in\epsfbox{Fig1.ps}}
%\bigskip\bigskip
%\caption{The linear-response conductance as a function of tunneling amplitude
%and temperature}
%\end{figure}
Fig. 1
(a) Transmission and (b) topographic images taken simultaneously 
on a 24$^{\circ}$ SrTiO$_{3}$ bicrystal.  The grayscales represent 30\% optical 
contrast in (a) and 30 \AA\ height difference in (b).  Most of the features 
in the topographic image are small debris on the surface.  The topographic 
image shows two small pits which correspond to the dark spots in the 
transmission image on the upper part of the image.  In general, when we 
observe a pit near the fusion boundary, there is always a corresponding 
dark spot in the transmission image.  However, the reverse is not the case; we 
observe far more dark spots in the NSOM transmission images than pits on 
the topographic images. 
\\
Fig. 2
3 $\mu$m x 3 $\mu$m (a) Transmission and (a) magnitude of linear 
retardance ($\phi$) image for an individual defect on a 24$^{\circ}$ 
SrTiO$_{3}$ bicrystal.  
The grayscales represent 10\% contrast in (a) and 0.05 radians in (b).  
The linecuts perpendicular to the fusion boundary indicated on (a) and (b) 
are shown in (c) and (d) respectively.  
\\
Fig. 3
Maximum $\phi$ 
for an individual defect vs. maximum optical contrast for 21 defects on 
4 SrTiO$_{3}$ bicrystals.  Different symbols are used based on the 
bicrystal on which they are found and the objective used for light 
collection.  A separate symbol is also used to denote two points 
which represent the same defect, but imaged with two different NSOM tips.
\\
Fig. 4
(a) Transmission and (b) $\phi$ image of the 
same area on the 36$^{\circ}$ bicrystal.  Two defects, marked with arrows, 
which look similar in the transmission image and have similar optical 
contrast but have very different retardance patterns and maximum 
$\phi$ values.  The lower defect has 40\% optical contrast and $\phi = 0.084$ 
radians, while the upper defect has 33\% optical 
contrast and is not  distinguishable in the $\phi$ image.


\begin{references}
\bibitem{36deg} The actual angle between the two sides of 
a 36$^{\circ}$ bicrystal is 36.8$^{\circ}$.
\bibitem{deltaI} Optical contrast is defined here as change in 
PMT current at the dark spot from the PMT current away 
from the bicrystal boundary divided by the PMT current away from the boundary 
$(\Delta I / I_{avg})$.
\bibitem{index} In addition to 670 nm, we performed 
the transmission NSOM measurements using laser light at 568 nm, 514.5 nm, 
and 488 nm.  The contrast increases with reducing wavelength, in 
agreement with the wavelength dependence of SrTiO$_{3}$ refractive index.  
The dependence of transmission contrast using different objectives 
also confirm that the observed contrast is a index not an absorption phenomenon.
\bibitem{expsetup} 670 nm laser light passes first through a 
linear polarizer (oriented at 90$^{\circ}$), and then through a 
photoelastic modulator (PEM), and finally through a quarter wave plate.  
The PEM introduces a sinusoidally time varying phase shift $\delta_0 \sin (2 \pi ft)$ 
into the +45$^{\circ}$ polarization component, where the modulation frequency 
$f$ is the resonant frequency (50 kHz) of the PEM quartz element.  
The emerging light has a polarization state which is varying in ellipticity at 
$f$ and has its principle axis at -90$^{\circ}$.  The quarter wave plate 
(oriented at 0$^{\circ}$) transforms this elliptically polarized light 
into linearly polarized light with its orientation varying sinusoidally 
at the modulation frequency.  This polarized light is then coupled into a 
single-mode optical fiber leading to the NSOM tip.  The light emitted from 
the tip aperture and transmitted through the sample is collected by a conventional 
microscope objective lens.  The collected light then passes through a 
circular polarization analyzer which transforms polarization state 
variation into intensity variation before reaching a photomultiplier tube 
(PMT).  The PMT signal therefore contains information about the birefringence 
of the sample and the overall transmittance of the sample.
\bibitem{objective} Optical contrast is measured using a 100x 0.7 NA objective.
\end{references}
\end{document}